\documentclass{ckm}                 

\confname{Workshop on the CKM Unitarity Triangle, IPPP Durham, April
  2003}

\title{Experimental review on moment analyses}

\author{Marta Calvi
\address{Dep. of Physics, University of Milano 
Bicocca and INFN, piazza della Scienza 3, I-20126 Milano, Italy}
}

\begin{document}

\begin{abstract}
Moments of the photon energy spectrum in $B\rightarrow X_s\gamma$ 
decays, of the hadronic mass spectrum and of the lepton 
energy spectrum in $B\rightarrow X_c\ell\nu$ decays are sensitive  
to the masses of the heavy quarks as well as to the 
non-perturbative parameters of the heavy quark expansion.
Several measurements have been performed both at the 
$\Upsilon(4S)$ resonance and at $Z^0$ center of mass energies.
They provide constraints on the non-perturbative parameters, give a test of 
the consistency of the theoretical predictions and of 
the underlying assumptions and allow to reduce the uncertainties
in the extraction of $|V_{cb}|$.
\vspace{1pc}
\end{abstract}

\maketitle

\section{Introduction}

The Operation Product Expansion is a powerful tool to study the dynamics of 
heavy flavour hadrons and represents a basis for 
extracting  the $|V_{cb}|$  element of the CKM mixing 
matrix from inclusive semileptonic B decays. 
In this framework inclusive observables are expressed in terms of 
quark masses
and non-perturbative effects are described by expectation values of 
heavy quark operators.
\par\noindent
As an example, when a double expansion in $\alpha_s$ and $1/m_B$ is used, 
the non-perturbative parameter introduced at first order 
is $\bar \Lambda$, which is related to the energy of the light degrees of 
freedom inside the heavy meson.
At second order $\lambda_1$ and $\lambda_2$ are introduced, which are
related to the kinetic energy of the b quark inside the meson and  
to the chromo-magnetic coupling of the b spin to the light degrees of freedom,
respectively. 
At third order other six parameters appear: 
$\rho_1,\rho_2,{\cal T}_1,{\cal T}_2,{\cal T}_3$ and ${\cal T}_4$.
Only $\lambda_2$ is relatively well known, from the $B^*-B$ meson mass 
difference.

In the determination of $|V_{cb}|$ from the inclusive semileptonic decay 
width 
\begin{center}
$\Gamma(b\rightarrow c \ell\nu) = |V_{cb}|^2 \gamma_{th}=
{BR(b\rightarrow c \ell\nu)}/{\tau_b}$
\end{center}
the uncertainties on the values of the non-perturbative parameters and on the 
assumptions underlaying the theoretical prediction, notably quark-hadron 
duality, provided  until recently the dominant error 
contribution~\cite{PDG02}. 
The current experimental uncertainties on the measurements 
of the semileptonic branching ratio and of the B lifetime give
a contribution of only about 1\%.

However other inclusive variables like
the moments of the photon energy spectrum in $B\rightarrow X_s\gamma$ 
decays and moments of the hadronic mass spectrum and of the lepton 
energy spectrum in $B\rightarrow X_c\ell\nu$ decays 
are sensitive to the masses of the heavy quarks and  to the 
non-perturbative parameters of the heavy quark expansion.
Measurements of spectral moments give constraints on the non-perturbative 
parameters and allow to reduce the uncertainties in the
extraction of $|V_{cb}|$~\cite{CKMyellow2002}.
Moreover, the comparison of results obtained from different measurements 
provides a test of the  consistency of the theoretical predictions and of 
the underlying assumptions.

The first measurements of spectral moments have been performed at the 
$\Upsilon(4S)$ by the CLEO Collaboration.
New preliminary results have been presented in summer 2002 by BABAR and DELPHI.
Moments of second and third order have been measured, having sensitivities to 
several parameters of the heavy quark expansion. 
For experimental reasons, measurements  performed up to now at the 
$\Upsilon(4S)$ require a minimum value of the lepton energy of about 1.5 GeV  
and thus have to compare to theoretical calculations which are also
restricted to a truncated lepton spectrum.

\section{CLEO}
\begin{figure}[ht]
\hbox to\hsize{\hss
\includegraphics[width=\hsize,height=7cm]{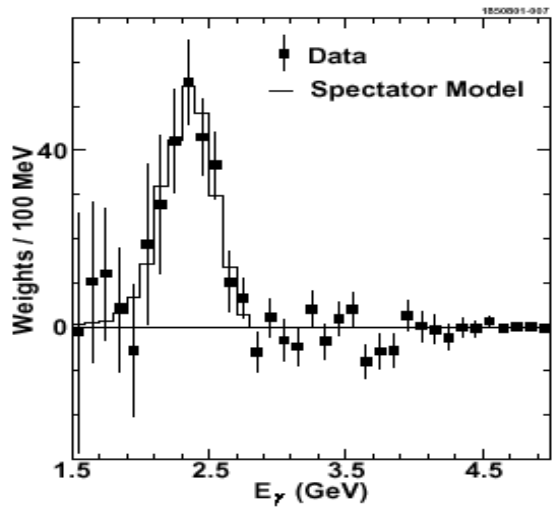}
\hss}
\caption{CLEO. Photon energy spectrum in $B\rightarrow X_s\gamma$ decays, 
in the laboratory frame, after 
background subtraction~\cite{cleo_Eg}.}
\label{fig:CLEO_Egamma}
\end{figure}
The photon energy spectrum in $B\rightarrow X_s\gamma$ decays has been 
studied with 9.1 $fb^{-1}$ of data collected at the $\Upsilon(4S)$ resonance
~\cite{cleo_Eg} with the CLEO detector. 
The large background of high energy photons from continuum processes has
been suppressed by using several techniques and the remaining part has been 
estimated with Monte Carlo and subtracted. 
The photon spectrum is shown Fig.~\ref{fig:CLEO_Egamma}.
After correcting for the experimental resolution, efficiency and smearing 
due to the B momentum, the first and second moment, for $E_{\gamma}>2$ GeV,
have been measured to be:
\begin{center}
$\langle E_{\gamma}\rangle =(3.346 \pm 0.032 \pm 0.011 )~GeV$\\
$\langle (E_{\gamma}-\langle E_{\gamma}\rangle)^2 \rangle
=(0.0226 \pm 0.0066 \pm 0.0020 )~GeV^2$\\
\end{center}

\begin{figure}[hb]
\hbox to\hsize{\hss
\includegraphics[width=\hsize,height=7cm]{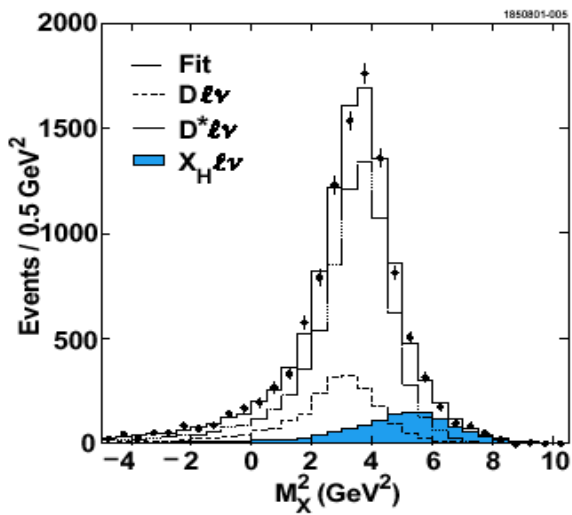}
\hss}
\caption{CLEO. Measured hadronic mass distribution for background 
corrected data. Different contributions in the Monte Carlo are also shown
~\cite{cleo_Mx}.}
\label{fig:CLEO_Mx}
\end{figure}

The hadronic mass spectrum has been  reconstructed in 
$B\rightarrow X_c\ell\nu$ decays using 3.2 $fb^{-1}$ of data
collected at the $\Upsilon(4S)$ resonance~\cite{cleo_Mx}.
The hermeticity of the CLEO detector has been exploited in reconstructing 
the neutrino by using momentum conservation of the entire event. 
Once the $B$ momentum, which is small ($\sim$ 300 MeV/$c$), is set
to zero, the mass of the hadronic system  
is determined from the lepton and neutrino momentum vectors alone as:
$\tilde{M}_X^2=M_B^2+M_{\ell\nu}^2-2E_BE_{\ell\nu}$.
Fig.~\ref{fig:CLEO_Mx} shows the $\tilde{M}_X^2$ distribution, for 
background-corrected data.
With the requirement $p_{\ell}>1.5$ GeV/$c$ the first two moments of 
the hadronic mass spectrum have been measured to be:
\begin{center}
$\langle M_X^2-m_{\bar D}^2\rangle=(0.251\pm0.023\pm0.062)~GeV^2$\\
$\langle(M_X^2-m_{\bar D}^2)^2\rangle=(0.576\pm0.048\pm0.163)~GeV^4$\\
\end{center}
where $m_{\bar D}$ is the spin-averaged D meson mass 
$m_{\bar D}=(M_D+3M{_D^*})/4$.

Using for the first moments 
the expressions  given in refs.~\cite{Ligeti-Luke} and \cite{Falk-Luke}, 
which are up to $1/m_B^3$ 
and $\alpha_s^2\beta_0$ order, 
the following constraints on $\bar{\Lambda}$ and $\lambda_1$ have been derived:
\begin{center}
$\bar{\Lambda}=(0.35 \pm 0.07 \pm 0.10 )~GeV$\\
$\lambda_1 =(-0.236 \pm 0.071 \pm0.078  )~GeV^2$
\end{center}
where the first uncertainty is the experimental one on the two moments
and the second is from the theoretical expressions.

\begin{figure}
\hbox to\hsize{\hss
\includegraphics[width=\hsize,height=7cm]{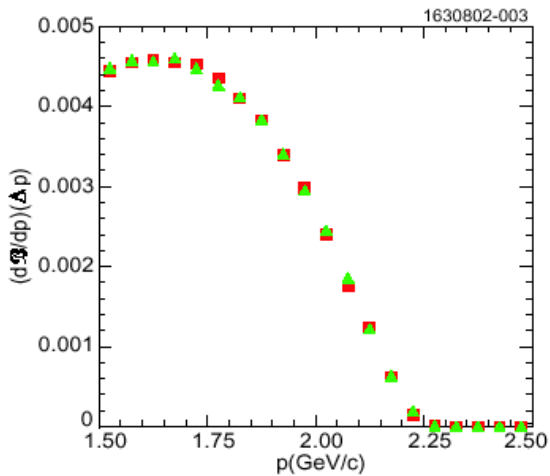}
\hss}
\caption{CLEO. Corrected electron (triangles) and muon (squares) energy 
spectra, in the B rest frame~\cite{cleo_Lept}.}
\label{fig:Cleo_lepton}
\end{figure}

New results on leptons have been recently published in ref.~\cite{cleo_Lept}.
The momentum spectra of electron and muon 
have been analysed using 3.1 $fb^{-1}$ of data collected at the $\Upsilon(4S)$
resonance.  A minimum momentum of  $p_{\ell}>1.5$ GeV/$c$ has been required,
 in order to ensure good efficiency and to reduce the contamination due 
to secondary leptons. In Fig.~\ref{fig:Cleo_lepton} the  electron and muon 
spectra are shown, after background subtraction.

\begin{figure}
\hbox to\hsize{\hss
\includegraphics[width=\hsize,height=8cm]{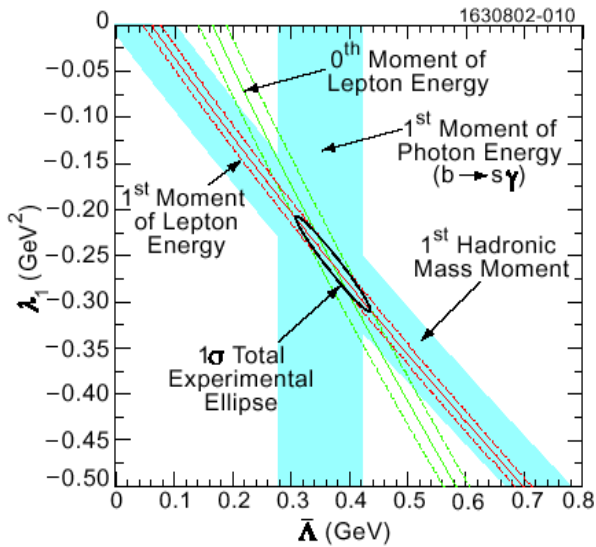}
\hss}
\caption{Constraints on the $\bar{\Lambda}$ and $\lambda_1$ parameters obtained
from CLEO measurements. The $\Delta\chi^2=1$ contour for the total 
experimental 
uncertainty is shown for the measurements of $R_0$ and $R_1$ moments.
The bands corresponding to the measurements of the first moment of photon 
energy spectrum and hadronic mass spectrum, respectively, are also shown.}
\label{fig:Cleo_param}
\end{figure}

The measurement of two moments of truncated lepton spectra, $R_0$ and $R_1$, 
have been used to constrain $\bar{\Lambda}$ and $\lambda_1$ \cite{Gremm} 
\begin{center}
$R_0=~0.6187 \pm 0.0014 \pm 0.0016$\\
$R_1=(~1.7810 \pm 0.0007 \pm 0.0009)$ GeV.
\end{center}
where
$$R_0={ {  {\int_{1.7GeV}} d\Gamma_{sl}/dE_{\ell} } 
\over  {  {\int_{1.5GeV}} d\Gamma_{sl}/dE_{\ell} } } ~~~~~~~
R_1={ {  {\int_{1.5GeV}} E_{\ell} d\Gamma_{sl}/dE_{\ell} } 
\over  {  {\int_{1.5GeV}} d\Gamma_{sl}/dE_{\ell} } }.$$
The dependence of $R_0$ and  $R_1$ on the parameters $\bar{\Lambda}$ and 
$\lambda_1$ is shown in Fig.~\ref{fig:Cleo_param}. The allowed region 
corresponds to the values:
\begin{center}
$\bar{\Lambda}=(~0.39 \pm 0.03 \pm 0.06 \pm 0.12)~GeV$\\
$\lambda_1 =(-0.25 \pm 0.02 \pm0.05 \pm 0.14  )~GeV^2$
\end{center}
where the quoted uncertainties are in order statistical, systematic and 
theoretical. The derived constraints are
in good agreement with those determined from the first moments of
photon energy spectrum and of hadronic mass spectrum measurements.
Using as input $\Gamma_{sl}^{exp}=(0.43\pm0.01)\times 10^{-10} MeV$ 
the following value of $|V_{cb}|$ has been obtained:
\begin{center}
$|V_{cb}|=(40.8\pm0.5\pm0.4\pm0.9) \times 10^{-3}$
\end{center}
where the first uncertainty is experimental, the
second is from the uncertainties in the non-perturbative  parameters
$\bar{\Lambda}$ and $\lambda_1$ and the third is the theoretical 
uncertainty determined by varying the input parameters within their 
respective errors. The uncertainty related to the truncation of the 
perturbative series is not included. 

\section{BABAR}

\begin{figure}
\hbox to\hsize{\hss
\includegraphics[width=\hsize,height=7cm]{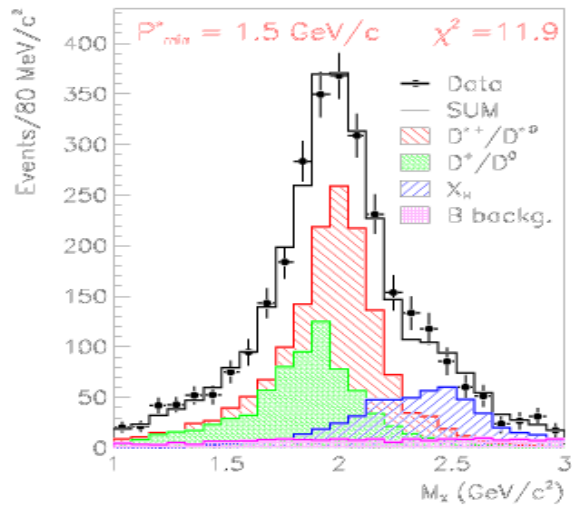}
\hss}
\caption{BABAR. Hadronic mass distribution for $p_{\ell}>1.5$ GeV/$c$.}
\label{fig:Babar_Mx}
\end{figure}

A preliminary measurement of the spectrum of the hadronic mass in
$B\rightarrow X_c\ell\nu$ decays has been 
presented by the BABAR Collaboration,
based on 51 $fb^{-1}$ of  data collected at the $\Upsilon(4S)$ resonance
~\cite{babar_Mx}.
Profiting of the high statistics, one of the B mesons ($B_{reco}$) is 
fully reconstructed in a hadronic channel and
the semileptonic decay of the other B is studied. The mass of the hadronic 
system is  reconstructed by summing all tracks in the event except the lepton 
and $B_{reco}$ and the mass resolution is improved by performing a
2-C kinematic fit to the whole event that imposes four-momentum  conservation. 
The mass distribution is shown in Fig.~\ref{fig:Babar_Mx}.
The mass distribution is fitted to the sum of four contributions 
accounting for decays in D$^*$ and D  mesons, in higher mass states 
$X_H$, including both resonant D$^{**}$ and non resonant D$^{*}\pi$ states, 
and background.
Monte Carlo simulation is used to derive the shape of the signal and 
background contributions.
A minimum momentum has been required for the lepton, which has been varied
 in the range 0.9 to 1.5 GeV/$c$. 
The contributions measured from the fit ($f_{D^*}$, $f_{D}$ and $f_{X_H}$)
are used to determine the first moment of hadronic mass spectrum as:
\begin{center}
$\langle m_X^2 - m_{\bar D}^2 \rangle = $\\
$f_{D^*}(m_{D^*}^2 - m_{\bar D}^2) + f_{D}(m_{D}^2 - m_{\bar D}^2) + 
f_{X_H}\langle M_{X_H}^2 - m_{\bar D}^2\rangle$
\end{center}
With the result obtained for $p_{\ell}>1.5$ GeV/$c$, and
using as a constraint on $\bar{\Lambda}$ the one derived by CLEO from 
$B\rightarrow X_s\gamma$ spectra $\bar{\Lambda}=(0.35 \pm 0.13)~GeV$,
$\lambda_1$ has been determined to be:
\begin{center}
$\lambda_1 =(-0.17 \pm 0.06 \pm0.07  )~GeV^2$ 
\end{center}
in good agreement with the $\lambda_1$ value derived by CLEO.
However the dependence of the first hadronic mass moment on $p^{min}_{\ell}$ 
has been found to be quite steeper than what is predicted by theory
\cite{Falk-Luke} and a fit to the data at different $p^{min}_{\ell}$ with
$\lambda_1$ and $\bar{\Lambda}$ as free parameters gives results incompatible 
with the previous ones.
A new analysis in under-way on BABAR data which will give more insight into
this question.

\section{DELPHI}

The first measurement of moments in $b$ hadron semileptonic decays at 
$Z^0$ center of mass energies has been performed by the DELPHI Collaboration.
The main advantage at the $Z^0$ pole is the large boost
 acquired by the b quark ($E_B\sim30$ GeV) which gives access in the 
laboratory frame to the low region of the lepton energy spectrum.
This makes these results both easier to interpret and complementary to 
those obtained at the $\Upsilon(4S)$.
The challenge in this case is the complete reconstruction of the B system.

Moments of  lepton (electrons and muons) energy spectrum have been 
measured in inclusive b-hadron
semileptonic decays~\cite{DELPHI_lept}.   
Secondary vertices have been reconstructed using an 
iterative procedure. The B energy has been determined adding to the 
energy at the  charm vertex the lepton energy and the neutrino energy, 
evaluated from the event missing energy. 
The B direction has been estimated from both the 
reconstructed B momentum and the B decay flight direction
and leptons have been boosted in the B rest frame. 
After unfolding the resolution smearing, the first, second and 
third moments have been determined to be:
\begin{center}
$\langle E_\ell \rangle = (1.383 \pm 0.012 \pm 0.008 )~GeV$\\
$\langle (E_\ell-\langle E_\ell\rangle)^2\rangle = (0.192 \pm 0.005 
\pm 0.010 )~GeV^2$\\
$\langle (E_\ell-\langle E_\ell\rangle)^3\rangle = (-0.029 \pm 0.005 
\pm 0.005 )~GeV^3$\\
\end{center}
For the measurement of hadronic mass moments  
${\rm \bar{B^0_d}}\rightarrow {\rm D}^{**}\ell\bar{\nu}$ events have been
studied ~\cite{DELPHI_had}.  
D$^{**}$ events have been  reconstructed in the three channels:
D$^0 \pi^+$, D$^+\pi^-$, D$^{*+}\pi^-$  
with the D$^0$, D$^+$ and D$^{*+}$ meson decays fully reconstructed.
Leptons have been required to have a momentum greater than 2~GeV/$c$ in the
laboratory frame. 
The separation of the signal from the background has been achieved by means
of a discriminant variable based on the topological properties of the 
secondary vertex.

\begin{figure}
\hbox to\hsize{\hss
\includegraphics[width=\hsize,height=7cm]{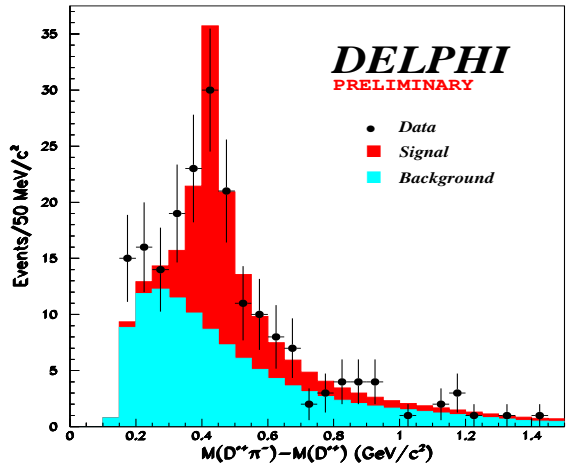}
\hss}
\caption{\sl DELPHI. $\Delta M$ distributions in the D$^{*+}\pi^-$ 
reconstructed decay channel.}
\label{fig:DELPHI_deltam}
\end{figure}

A fit to the variable $\Delta M=m({\rm D}^{(*)}\pi)-m({\rm D}^{(*)})$ 
has been performed, with contributions of narrow and broad resonant states  
D$_0^{*+}$, D$_1^{*+}$, D$_1^{+}$ and D$_2^{+}$  as well as non resonant 
D$\pi$ states free in the fit. Constraints on available measurements on 
narrow states have also been applied.
The total rate for D$^{**}$ production resulting from the fit
\begin{center}
$BR({\rm \bar{B^0_d}}\rightarrow {\rm D}^{**}\ell\bar{\nu})
=(2.6\pm0.5\pm0.6)\%$ 
\end{center}
is well compatible with previous measurements.
The $\Delta M$ distribution in one of the reconstructed decay channels
 is shown in Fig.~\ref{fig:DELPHI_deltam}.
From the fitted mass distributions the first three moments of the 
D$^{**}$ mass distribution have been evaluated. 
The relation:
\begin{center}
$ \langle m_X^n \rangle = p_D m_D^n +  p_{D^*} m_{D^*}^n +
                        p_{D^{**}} \langle m_{D^{**}}^n \rangle $
\end{center}
has been used to add the $b\rightarrow {\rm D}\ell\bar{\nu}$ and  
the $b\rightarrow {\rm D}^*\ell\bar{\nu}$ contributions, where
$p_{D}$ and $p_{D^*}$ are the relative branching fractions derived  
from published results and $p_{D^{**}}$ is obtained by imposing the 
constraint $p_D +  p_{D^*} + p_{D^{**}} =1$ and using the above measurement.

The following preliminary results have been obtained:
\begin{center}
$\langle m_X^2 - m_{\bar D}^2 \rangle = (0.534 \pm 0.041
                                          \pm 0.074 ) ~GeV$\\
$\langle (m_X^2 - \langle m_X^2\rangle )^2 \rangle = (1.23 \pm 
                                                0.16 \pm 0.15 ) ~GeV^2$\\
$\langle (m_X^2 - \langle m_X^2\rangle )^3 \rangle = (2.97 \pm 
                                                0.67 \pm 0.48 ) ~GeV^3$\\
\end{center}
\par\noindent
where the first uncertainty is statistic and the second is systematic.

From the measured spectral moments constraints on the non-perturbative 
parameters of the OPE have been obtained~\cite{Batt}. 
The formalism based on low scale running masses ~\cite{Bigi}
which does not rely on a $1/m_c$ expansion has been used.
Here $m_b(\mu)$ and $m_c(\mu)$ are independent parameters,
$\mu_{\pi}^2$ has a similar meaning than $\lambda_1$,
and the only parameters appearing at order $1/m_b^3$ are $\rho_D^3$ and 
$\rho_{LS}^3$.  Expressions for non-truncated lepton spectrum have been used
and higher moments have been included to gain sensitivity on  
$1/m_b^3$ parameters.
A multi-parameter $\chi^2$ fit has been performed to the first three 
moments of the hadronic mass spectrum and lepton energy spectrum with
the following results: 
\begin{center}
$m_b(1 GeV)=(~4.59 \pm 0.08 \pm 0.01)~GeV$\\
$m_c(1 GeV)=(~1.13 \pm 0.13 \pm 0.03)~GeV$\\
$\mu_{\pi}^2(1 GeV)=(0.31\pm0.07\pm0.02)~GeV^2$\\
$\rho_D^3(1 GeV)=(0.05\pm0.04\pm0.01)~GeV^3$\\
\end{center}
\par
\begin{figure*}[ht]
\hbox to\hsize{\hss
\includegraphics[width=\hsize]{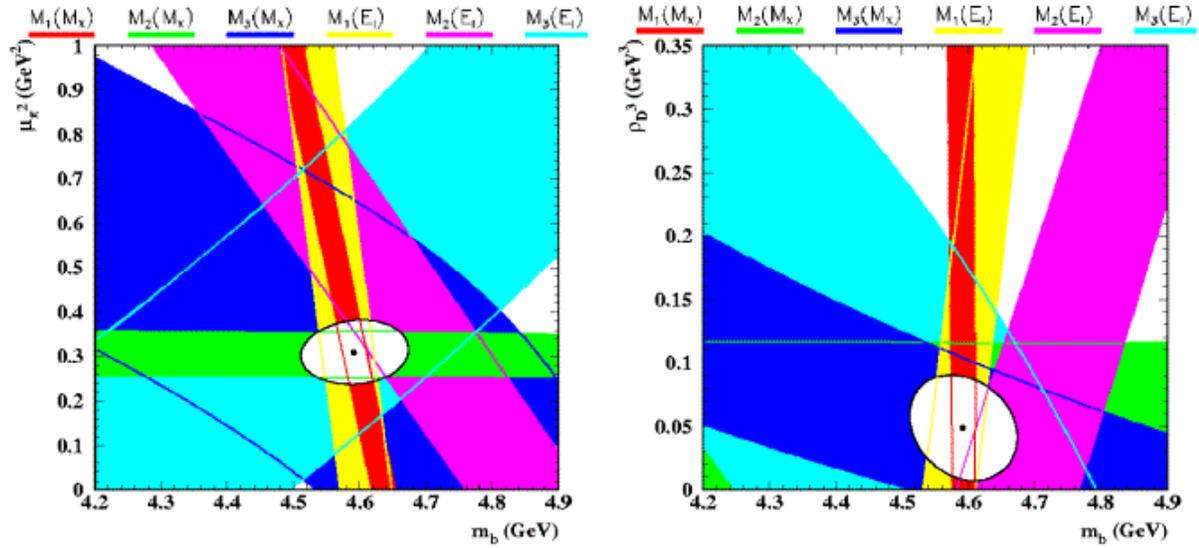}
\hss}
\caption{\sl The constraints on the 
$\mu_{\pi}^2 - m_b$ (left)  and $\rho_D^3 - m_b$ (right) plane 
obtained from the combination of the first three moments of the lepton energy 
spectrum and hadronic mass spectrum.
The bands correspond to the total experimental accuracy on each moment,
and are given by keeping all the other parameters at their central values.
The ellipses represent the $1 \sigma$ contour.}
\label{fig:fit_mb}
\end{figure*}

\noindent
A good consistency between all measurements has been obtained, demonstrating
no need to introduce higher order 
terms to establish agreement with data, within the present accuracy.
Fig.~\ref{fig:fit_mb} shows the constraints extracted in the 
$\mu_{\pi}^2 - m_b$  and $\rho_D^3 - m_b$ planes. 

The fit has been repeated also using the the pole mass 
formalism~\cite{Falk-Luke} obtaining: 
\vspace*{-.3cm}
\begin{center}
$\bar{\Lambda}=(~0.40\pm 0.10\pm0.02 )~GeV$\\
$\lambda_1 =(-0.15 \pm 0.07 \pm0.03 )~GeV^2$\\
$\rho_1=(-0.01\pm0.03\pm0.01)~GeV^3$\\
$\rho_2=(0.03\pm0.03\pm0.01)~GeV^3$\\
\end{center}
\par\noindent
compatible with CLEO results.
The first uncertainties are from the fit, the second are systematic, coming
from the variation of the residual parameters which have been fixed in the fit and from missing terms in the expansions.

The value of $|V_{cb}|$ obtained from the inclusive semileptonic decay 
width depends on the OPE parameters extracted above \cite{Kolya}. 
An approximate formula which displays the dependence on the different 
parameters is the following:
\vspace*{-.2cm}
\begin{center}
$|V_{cb}|= |V_{cb}|_0~  \{ 1 + $\\
$-0.65 ~[m_b(1)-4.6~GeV] + 0.40~[m_c(1)-1.15~GeV]$\\
$+0.01 ~[\mu_{\pi}^2(1)-0.4~GeV^2] + 0.10 ~[\rho_D^3-0.12~GeV^3]$ \\
$+0.06 ~[\mu_{G}^2(1)-0.35~GeV^2] + 0.01 ~[\rho_{LS}^3+0.17~GeV^3] \}$\\
\end{center}
Using the the world average value of the semileptonic width
$\Gamma_{sl}^{exp}=(0.434\pm0.008)\times 10^{-10} MeV$
the following value of $|V_{cb}|$ has been obtained:
\vspace*{-.2cm}
\begin{center}
$|V_{cb}|=(41.6\pm0.4\pm0.6\pm0.4) \times 10^{-3}$
\end{center}
where the first uncertainty is from the semileptonic width, the 
second is from the uncertainties in the non-perturbative  parameters
determined from the fit and the third is from 
the variation of $\alpha_s$ scale.

A similar analysis has been performed in~\cite{Bauer-Ligeti} using CLEO and 
DELPHI measurements and consistent results have been obtained.

\section{Conclusions}
Measurements of moments of the photon energy spectrum in 
$B\rightarrow X_s\gamma$ 
decays, of the hadronic mass spectrum and of the lepton 
energy spectrum in $B\rightarrow X_c\ell\nu$ decays 
have been performed both at the $\Upsilon(4S)$ and at $Z^0$ center of mass 
energies. New results will also come from the B factories in the near future.
Good agreement has been found between the values of the non-perturbative 
parameters of the heavy quark expansions derived from different measurements
and no hint of a possible quark-hadron duality violation effect has emerged, 
within the present accuracy.
As a consequence the uncertainties in the extraction of $|V_{cb}|$ from 
inclusive semileptonic with has been much reduced.

\end{document}